\documentclass{article}

\usepackage{arxiv}

\usepackage[utf8]{inputenc} % allow utf-8 input
\usepackage[T1]{fontenc}    % use 8-bit T1 fonts
\usepackage{hyperref}       % hyperlinks
\usepackage{url}            % simple URL typesetting
\usepackage{booktabs}       % professional-quality tables
\usepackage{amsfonts}       % blackboard math symbols
\usepackage{nicefrac}       % compact symbols for 1/2, etc.
\usepackage{microtype}      % microtypography
\usepackage{lipsum}
\usepackage{graphicx}
\usepackage{xcolor}
\usepackage{array, makecell}
\graphicspath{ {./images/} }

\title{Quality of Automatic Speech Recognition - Polish Language case study - from Wav2Vec to Scribe ElevenLabs}

\author{
 Marcin Pietroń \\
  Institute of Electronics\\
  AGH, Krakow \\
  \texttt{pietron@agh.edu.pl} \\
   \And
 Szymon Piórkowski\\
  Coraz Zdrowiej, Krakow \\
  \texttt{s.piorkowski@corazzdrowiej.pl} \\
  \And
 Kamil Faber \\
  Department of Computer Science\\
  AGH, Krakow \\
  \texttt{kfaber@agh.edu.pl} \\
  \And
  Dominik Żurek \\
  Department of Computer Science\\
  AGH, Krakow \\
  \texttt{dzurek@agh.edu.pl} \\
  \And
  Michał Karwatowski \\
  Institute of Electronics\\
  AGH, Krakow \\
  \texttt{mkarwat@agh.edu.pl} \\
  \And
  Jerzy Duda \\
  Department of Management\\
  AGH, Krakow \\
  \texttt{jduda@agh.edu.pl} \\
  \And
  Hubert Zieliński \\
  Coraz Zdrowiej, Krakow \\
  \texttt{h.zielinski@corazzdrowiej.pl} \\
  \And
  Piotr Lipnicki \\
  Coraz Zdrowiej, Krakow \\
  \texttt{p.lipnicki@corazzdrowiej.pl} \\
  \And
  Mikołaj Leszczuk \\
  Department of Telecommunication\\
  AGH, Krakow \\
  \texttt{mikolaj.leszczuk@agh.edu.pl} \\
  \And
}

\begin{document}
\maketitle
\begin{abstract}
This article concerns comparative studies on the Automatic Speech Recognition (ASR) model incorporated with the Large Language Model (LLM) used for medical interviews. The proposed solution is tested on polish language benchmarks and dataset with medical interviews.
%In such systems, recognition is based on Mel Frequency Cepstral Coefficients (MFCC) acoustic features and spectrograms.
The latest ASR technologies are based on convolutional neural networks (CNNs), recurrent neural networks (RNNs) and Transformers. Most of them work as end-to-end solutions. The presented approach in the case of the Whisper model shows a two-stage solution with End-To-End ASR and LLM working together in a pipeline. The ASR output is an input for LLM. The LLM is a component by which the output from ASR is corrected and improved. 
%The prompt techniques are performed to force the LLM to find wrong words or phrases. 
%The LLM used in presented simulation is Bielik, a large language model fine tuned for polish language. 
Comparative studies for automatic recognition of the Polish language between modern End-To-End deep learning architectures and the ASR hybrid model were performed. 
%The differences between conventional architectures and ASR DNN End-To-End (E2E) models are discussed. 
The medical interview tests were performed with two state-of-the-art ASR models: OpenAI Whisper incorporated with LLM and Scribe ElevenLabs. Additionally, the results were compared with five more end-to-end models
(QuartzNet, FastConformer, Wav2Vec 2.0 XLSR and ESPnet Model Zoo) on Mozilla Common Voice and VoxPopuli databases. Tests were conducted for clean audio signal, signal with bandwidth limitation, and degraded. The tested models were evaluated on the basis of Word Error Rate (WER) and Character Error Rate (CER). The results show that the Whisper model performs by far the best among the open-source models. ElevenLabs' Scribe model, on the other hand, performs best for Polish on both general benchmark and medical data.

%The results show that LLM can drastically improve the output of the end-to-end ASR. To our knowledge, this is the first attempt to use an advanced language model in a speech-to-text task for the Polish language.
\end{abstract}

% keywords can be removed

\section{Introduction}
%\textcolor{%red}{Large Language Models as ASR fine tuners, whisper}
The ability to convert spoken language into written text, known as Automatic Speech Recognition (ASR), is an important element of the development of human-computer interaction. ASR systems have become integral components
in numerous applications, from voice assistants to transcription services, and have found applications in a wide array of languages. Due to its increasingly higher effectiveness, it is used in an increasing number of applications.
%However, for 
The ASR for the Polish language, due to its’ complicated structure and limited data resources, still requires further research and improvements. In recent years, the emergence of End-To-End (E2E) approaches based on deep neural networks (DNNs) has accelerated ASR research, and E2E DNN systems show promise in processing Polish speech. Additionally, new large language models create opportunities to improve transcription performance without additional costly and lengthy training.

%This paper presents preliminary tests of five E2E ASR models adapted for Polish language recognition, 
Firstly, this paper presents preliminary tests conducted on the Mozilla Common Voice , Multilingual LibriSpeech (MLS) and VoxPopuli (VP) databases. Models adapted for automatic speech recognition in Polish were tested, two models available in the NVIDIA NeMo toolkit: QuartzNet and FastConformer Transducer-CTC), Whisper (developed by OpenAI), ESPnet2, Wav2Vec 2.0 XLSR-53 model (developed by MetaAI). Whisper and ESPnet2 are multilingual general-purpose models with a language detection stage before recognition. The Wav2Vec 2.0 version is an additional fine-tuned version of the multilingual model (which originally also covered Polish), which can be used as an ASR model for Polish. QuartzNet and FastConformer models have been fine-tuned for Polish based on pre-trained English models.
The paper then shows how ASR results can be improved in a two-step process by using a large language model. The quantitative results of this configuration are presented in the Whisper model for standard benchmarks. However, popular benchmarks are characterized by certain limitations, such as relatively short context, short sentences, and a lack of complex phrases or specialized words. These models were also largely trained on diverse training sets, often primarily based on everyday language. This significantly complicates the process of model generalization to data with different contexts and specialized vocabulary. Therefore, the results of models tested on standard benchmarks may not be sufficiently objective. To accurately assess the usefulness of models for specialized applications, it is necessary to conduct tests on properly prepared data within a given domain.

For this reason, the aim of the second part of the work was to test a subset of the best models (Whisper and ElevenLabs) on real-world medical interviews. Tests were conducted on dozens of recordings of varying quality. The results demonstrate varying models' generalization abilities, which translates into their performance. The results support the hypothesis that differences in model performance on standard benchmarks do not translate into differences in performance on specialized data. In the final part, fine-tuning of the Whisper model (an open-source model) was performed to demonstrate the model's adaptability to new data.

%\textcolor{red}{docelowo dla systemu notatek medycznych, fig.2}

\section{Related works}

In the field of automatic speech recognition, there are two main approaches: Conventional (Fig. 1), based on three models: acoustic (AM), linguistic (LM) and pronunciation (PM), and End-To-End (E2E), based on an integrated deep neural model (DNN). In both solutions, the input is an audio speech signal, and the output results in a transcription, i.e. a textual notation of the content contained in the input signal. The both approaches are distinguished by the models used, the preparation of training data, training process, and post-processing.

In the Conventional ASR approach (Fig. 1), models are trained separately and exchange information with each other. Typically, such systems are based on the HMM (Hidden Markov Models) Acoustic Model and the n-gram Language Model and PM as a lexicon/pronunciation dictionary. The process of preparing training data for conventional ASR systems is, therefore, complex and time-consuming.
HMM consists of Hidden Markov chains and observed variables. In ASR, hidden states correspond to phonemes (graphical representation of speech sound) and observed variables represent to sound frames (acoustic features). Using HMM requires segmenting the speech signal into smaller parts and assigning them the corresponding graphical representations. In addition, the Conventional ASR Acoustic Model uses a wide range of acoustic features, such as formant frequencies, Perceptual Prediction Coefficients (PLP), power normalized cepstral coefficients (PNCC) or Mel Frequency Cepstrum Coefficient (MFCC). This means that training the Acoustic Model in Conventional ASR requires additional steps. Most of these steps are eliminated or partially eliminated in the E2E DNN approach. In E2E subsequent words are predicted based on previous words, and the meaning of the utterance is discovered based on the local context. Statistical approaches are appropriate for English, which is positional and has a specific sentence formation (e.g. subject-verb-object (SVO)). For Polish, which is inflected and has almost arbitrary sentence formation, coverage of both local and global context is required.

%The requirement of Polish to simultaneously capture the local and global demands a different approach to building ASR systems. 
Research is considering E2E ASR architectures based on an integrated deep neural network model. Such an architecture is simpler to train. The weakness of such an architecture is the huge computational power requirements, which are constantly increasing with the development of larger models. 
%and new types of neural networks. 
One of the first deep neural networks used in the E2E ASR approach was Recurrent Neural Networks (RNNs), which are suitable for the analysis of sequential data, and their prediction is not only based on input at a given time, but is also updated with previous predictions. 
%However, they require a data pre-segmentation. 
In RNNs, the network has access to the entire previous sequence, but covers more local context - for the last parts of the input sequence. % and the prediction [21].
Other deep neural networks used in E2E ASR architectures are convolutional networks (CNNs). CNNs were originally applied to image recognition, but can be implemented for speech recognition in combination with its graphical representation (spectrograms). The context of the time dependence of speech is covered in the relevant depth of CNNs. 
%It doesn’t precise require pre-segmentation of data (on phonemes and phoneme labelling during model fine-tuning). 
%The CNN network filter moves across the mel-spectrogram of whole utterance (e.g a word), so it doesn’t require to be divided into smaller segments. 
The latest deep neural networks with immense potential in ASR systems are Transformers. It uses a MultiHeaded Self-Attention mechanism covering both global and local contexts. They were originally adapted for the task of natural language processing (NLP), i.e. texts. The architectures of E2E ASR models using Transformers networks are enriched with CNN layers, which process features of spectrograms, and a CTC loss function or Transducer (which have their origin in RNNs). The most effective E2E ASR models, therefore, seem to be those built from layers of all the above-mentioned types of deep neural networks.
Such representatives include the ASR Whisper and ElevenLabs models \cite{LLM, ElevenLabs}.

%\textcolor{red}{Large Language Models - GPT, LLAMA, Bielik, RAGs}

\section{Task description and data construction}

%\textcolor{red}{dataset description}

\begin{figure}[t!]
\centering
\includegraphics[width=0.8\textwidth]{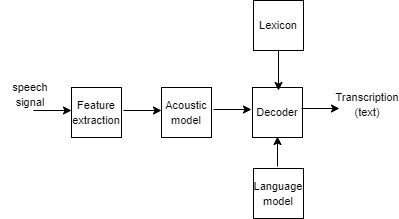}
\caption{ASR architecture}
\label{images}
\end{figure}

%To testing the E2E ASR architecture, suitable databases had to be prepared. 

Six open-source DNNs models were selected for the experiment, which are adapted to Polish speech recognition: QuartzNet, FastConformer, Wav2Vec 2.0, OpenAI Whisper, ESPnet2 and Scribe ElevenLabs. The QuartzNet model is fully convolutional. %with a BxR block architecture. 
Other models have an Encoder-Decoder architecture in which the Encoder converts the input audio signal into high-level representations and then the Decoder decodes the content and returns the most probable transcription.
The Whisper and Wav2Vec 2.0 models are OpenAI and MetaAI (formerly Facebook AI) projects, respectively. 
%The Whisper model is available on the Github [34] platform, and the Wav2Vec 2.0 XLS adapted to Polish language originates from the Hugging Face [35] platform (an AI community sharing database and model knowledge). 
%QuartzNet and FastConformer models are available in the NVIDIA NeMo Toolkit [4], and Model Zoo in the ESPnet toolkit [36]. 
The Whisper and ESPnet2 are general-purpose multilingual models which include Polish among their known languages. The described versions of the QuartzNet, FastConformer and Wav2Vec 2.0 models are fine-tuned to Polish based on pre-trained models recognizing speech in English. These characteristics of the selected models is as follows:

\begin{itemize}
\item QuartzNet: The QuartzNet model is a fully convolutional model (CNN) based on the Jasper model architecture - a deep time-delay neural network [39] (TDNN) consisting of blocks of layers of 1D CNNs. The model has a BxR architecture, where B is the number of blocks and R is the number of convolutional sub-blocks in a block. The QuartzNet model is distinguished from the pure Jasper version by separable splices and larger filters, making it perform close to Jasper, with an order of magnitude fewer parameters. 
%The model chosen for the experiment is a QuartzNet version fine-tuned to Polish, trained on MCV 6.0.
\item Conformer is a convolution augmented Transformer. In this model, the CNN layers are not just the initial feature processing layers (as in SpeechTransformer), but the Transformer block has been replaced by a Conformer with additional CNN layers behind the Multi-Head Self Attention. FastConformer is an optimized version of the Conformer model. The output of the encoder can be decoded using RNN-T or CTC loss (RNN-T by default). 
%The model was trained on the MCV 12.0, MLS and VP databases.
\item Wav2Vec 2.0:  model developed by MetaAI. It is based on CNNs and Transformers networks. The model is self-supervised, learning speech structure from raw audio. Wav2Vec 2.0 is a cross-linguistic approach to teach speech units common to several languages. The model does not require data labeling, forced alignment, or segmentation. The large Wav2Vec 2.0 model has been trained for 53 languages on the MLS, MCV and BABEL databases (a total of 56,000 hours of speech data, all of which include Polish). 
%The model used in the experiment is an additional fine-tuned version of Wav2Vec 2.0 XLSR-53 for Polish using MCV 6.1 training and validation parts [9]. 
\item OpenAI Whisper: Whisper is a general-purpose speech recognition model, has been trained on a large set of varied audio data, and can perform multilingual speech recognition, speech translation, and language identification. Using a command line, it is possible to determine the recognition language. When used in a Python implementation, the entire model is loaded first, and the first stage of recognition is language identification, followed by transcription. The Whisper architecture is based on the Speech-Transformer architecture. %Whisper model documentation includes results of evaluation based on Fluers [46] dataset, for 57 languages, including Polish (WER = 5.4\%).
\item ESPnet2: In ESPnet2, audio data is directly entered into the model as in all of the above models. ESPnet Model Zoo is a population of models trained on multiple datasets, both uni- and multilingual, available for public use. The Zoo model implemented in ESPnet2 consists of over 164 models, of which over 20 are intended solely for ASR purposes. Whereas most of those models were designed for English language recognition only, there is a limited choice of models that can be used for other languages. 
%Considering that Polish is a low-resource language, only “open li52” corpus [48] from ESPnet toolkit (containing 52 languages, including the MCV database) could actually be applied here. 
ESPnet2 Model Zoo is Transformer based and was trained on multilingual dataset containing tokens with Polish diacritic signs, letters with dashes, overdots and tails.

\item Scribe ElevenLabs Speech to Text model, is the world’s most accurate transcription model. Built to handle the unpredictability of real-world audio. Scribe transcribes speech in 99 languages, featuring word-level timestamps, speaker diarization, and audio-event tagging—all delivered in a structured response for seamless integration. Scribe makes ASR universally accessible and dramatically reduces errors in traditionally underserved languages such as Serbian, Cantonese, and Malayalam, where competing models often exceed 40\% word error rates.

\end{itemize}

\subsection{Standard transcription benchmarks}
For the first experiments, 2 multilingual databases containing the Polish language were selected. Six deep E2E ASR models based on both RNNs, CNNs and Transformers were selected for testing. %This Section describes the chosen data resources and models. 
Databases containing both recordings and reference texts are required to test selected E2E ASR models. The recordings are used for transcription, and the reference texts are used for model evaluation %(selected evaluation methods are described in Section III-C). 
(for the experiments, Polish parts from multilingual open-source databases were selected - Mozilla Common Voice (MCV), Multilingual LibriSpeech (MLS)):
\begin{itemize}
\item Mozilla Common Voice (MCV): database provided for 112 languages (including Polish) and covers MP3 recordings of speech, corresponding transcriptions (text) and metadata on age, gender and accent. The datasets for each language are divided into training (train), development (dev) and test sets. The Polish portion of MCV is constantly being expanded and currently contains 173 hours of speech recordings (163 hours verified)
%in 15.0 version [27]. The MCV version available at [28] was sampled for the research. 
%\item Multilingual LibriSpeech (MLS): database is a multilingual version of the LibriSpeech [29] database (only for English). MLS includes, in addition to English, 7 other languages (including Polish). It contains speech read from publicly
%available LibriVox [30] audiobooks and Project Gutenberg text data [31] (44,500 hours in English and a total of 6,000 hours in other languages). The dataset is divided into training, development and testing sets. The Polish set contains recordings in subsets of respectively: 103.65, 2.08 and 2.14 hours. 
%\item VoxPopuli (VP): database is a multilingual speech corpus (includes 23 languages, including Polish) that contains recordings from the European Parliament from 2009 to 2020. The database includes unlabeled (400,000 hours) and transcribed (1,800 hours for 16 languages, with Polish); speech-to-speech interpretation data and transcribed accented speech data [33]. The Polish part contains 21.2 thousand hours of non-transcribed recordings, including 111 hours for which transcription is available.
\item Multilingual LibriSpeech (MLS) dataset is a large multilingual corpus suitable for speech research. The dataset is derived from read audiobooks from LibriVox and consists of 8 languages: English, German, Dutch, Spanish, French, Italian, Portuguese, and Polish. Consists of train, dev and test sets for each language. Also, includes a small training set for limited supervision (10 h, 1 h, or 10 minutes of labeled speech).

\end{itemize}

\subsection{Medical interview transcription benchmarks}

Automatic transcription of medical interviews represents a particularly challenging and sensitive case for ASR systems. Compared to datasets such as CommonVoice or LibriSpeech, doctor–patient conversations are characterized by significant acoustic variability, overlapping speech, spontaneous phrasing, and frequent use of medical terminology. In addition, recordings often contain typical clinical environments—air-conditioning, electronic equipment, or distant microphones.

The medical data set used in our evaluation was constructed from real outpatient consultations recorded in Polish medical facilities. Each sample contains an audio file and its manually verified transcript. The corpus covers over 9 hours of speech distributed across approximately 260 dialogs, involving both physicians and patients. Individual recordings vary in duration, from short symptom descriptions (under 10 seconds) to complete anamneses that exceed several minutes.

The metadata extracted from the data set indicate a broad spectrum of medical contexts, including internal medicine, pediatrics, dermatology, and gynecology. Conversations were recorded under realistic acoustic conditions using consumer-grade microphones, which allowed us to reproduce real-world challenges, such as uneven loudness, interruptions, and overlapping speech. Each entry in the data set contains information about signal quality, average signal-to-noise ratio (SNR), speaker role (doctor/patient), and transcript completeness.

\section{System architecture}

\begin{figure}[t!]
\centering
\includegraphics[width=0.9\textwidth]{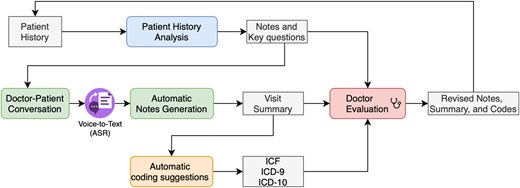}
\caption{CORA AI architecture}
\label{images}
\end{figure}

%\textcolor{red}{figure, whisper+LLM(Bielik)+prompting, figure, ElevenLabs, noisy data, halucynacje, medical data, model with RAG}

%This work tested ASR models for the Polish language. Simulations were performed using standard benchmarks and medical data. 
The collected test data in the form of medical notes and the test process are carried out using the system is described in Fig.2. It consists of the following modules:

\begin{itemize}
\item  Doctor-Patient Conversations - medical interview recording module 
\item  Automatic Notes Generation - medical notes generated by ASR model
\item  Visit Summary - module for generating summary from transcription
\item  Visit History - access to patient visit history 
\item  Automatic Coding suggestions - this module is responsible for disease code prediction
\item  Patient History and Patient History Analysis - the part of the system responsible for analyzing the patient's medical history
\item  Doctor Evaluation - evaluation of the generated data by a specialist
\item  ICF, ICD-9, and ICD-10 - acceptance of the final disease code
\end{itemize}

%The processing pipeline used in this study follows the classical ASR scheme, where the incoming speech signal is first transformed into acoustic features and then analysed by the acoustic model. The decoder integrates information from the language model and lexicon to produce the textual transcription. 

In our system, this transcription constitutes the input for the downstream pipeline that automatically generates structured visit notes, proposes diagnostic codes (ICF, ICD-9, ICD-10), and produces summaries for physician review. One of the most important elements of the system is the ASR model. It has the greatest impact on reducing the execution time of the entire pipeline. Secondly, it influences the quality of the subsequent summarization. Summarization (Visit Summary) is the second element of the system that requires a large deep learning model (LLM).

%\subsection{Scribe - ElevenLabs}

%Scribe is engineered for precision. In FLEURS  Common Voice benchmark tests across 99 languages, it consistently outperforms leading models like Gemini 2.0 Flash, Whisper Large V3 and Deepgram Nova-3. Whether it’s meeting summaries, movie subtitles, or even song lyrics, Scribe delivers the lowest automated transcription word error rate in Italian (98.7%), English (96.7%) and 97 other languages.

%\subsection{ASR}

%\subsection{ASR with RAG}

\section{Results on common benchmarks}

\begin{table}
\centering
\begin{tabular}{|c|c|}
\hline
\textbf{Model} & WER \\\hline
\textbf{ElevenLabs}     &    6.77\%    \\ \hline
\textbf{Whisper large}     &   5.05\%   \\ \hline
\textbf{Wav2Vec}     &    17.2\%    \\ \hline

\end{tabular}
\caption{Comparison between different ASR models - LibriSpeech}
\label{mls_results}
\end{table}

\begin{table}
\centering
\begin{tabular}{|c|c|c|c|c|c|c|c|c|c|}
\hline
\textbf{} & \makecell{\textbf{all} \\ \textbf{words}} & \makecell{\textbf{wrong} \\ \textbf{words}} & \makecell{\textbf{correct} \\ \textbf{sentences}} & \makecell{\textbf{wrong} \\ \textbf{sentences}}  & \makecell{\textbf{all} \\ \textbf{sentences}} & \textbf{WER}
& \textbf{F1 words} & \textbf{F1 sentences}\\ \hline
\textbf{ElevenLabs}     &    16 324     &  344  &      263         & 257 & 520 & 6.77 &   95.8  & 50.57           \\ \hline
\textbf{Whisper large}     &   17 140        &  757  &     172          &  17 034 &   520   &  5.05 &   95.87 & 33.07       \\ \hline
\textbf{Whisper medium}     &    17 084     &         1 061        &      116       & 404 &  520  & 6.84 &  93.92 & 22.3\\ \hline
\textbf{Whisper small}     &    17 071     &         1 810        &     43        &  477  &  520  & 11.92 &  89.49 & 8.26 \\ \hline

\end{tabular}
\caption{Comparison between different Whisper models and ElevenLabs for LibriSpeech}
\label{mls__results}
\end{table}

\begin{table}
\centering
\begin{tabular}{|c|c|c|c|c|c|c|c|c|c|}
\hline
\textbf{} & \makecell{\textbf{all} \\ \textbf{words}} & \makecell{\textbf{wrong} \\ \textbf{words}} & \makecell{\textbf{correct} \\ \textbf{sentences}} & \makecell{\textbf{wrong} \\ \textbf{sentences}}  & \makecell{\textbf{all} \\ \textbf{sentences}} & \textbf{WER}
& \textbf{F1 words} & \textbf{F1 sentences}\\ \hline
\textbf{ElevenLabs}     &    975 145     &  33 166  &        110 826          & 21 570 & 132 396 & 3.29 &   97.31  & 83.7           \\ \hline
\textbf{Whisper large}     &   975 145        &  59 487  &     94 867          &  37 529 &   132 396   &  6.3 &   94.54 & 71.65       \\ \hline
\textbf{Whisper medium}     &    975 145    &         104 303        &    75 756         &  56 640 &  132 396  &  11.65 & 89.76 & 57.21\\ \hline
\textbf{Whisper small}     &    975 145     &   167 827              &      54 108       & 78 288 &  132 396  & 19.28 & 83.04 & 40.86 \\ \hline

\end{tabular}
\caption{Comparison between different Whisper models and ElevenLabs for CommonVoice}
\label{cv_results}
\end{table}

\begin{table}
\centering
\begin{tabular}{|c|c|}
\hline
\textbf{Model} & WER \\\hline
\textbf{ElevenLabs}     &    3.29\%    \\ \hline
\textbf{Whisper large}     &   6.3\%   \\ \hline
\textbf{QuartzNet}     &    14\%    \\ \hline
\textbf{FastConf}     &    5.99\%      \\ \hline
\textbf{ESPnet2}     &    15.1\%     \\ \hline
\textbf{Wav2Vec}     &    9.8\%    \\ \hline

\end{tabular}
\caption{Comparison between different ASR models - Common Voice}
\label{cv__results}
\end{table}

%\textbf{With special characters}

%all words = 971 773 \\
%wrong words = 20 5065 \\
%correct sentenctes = 51 993 \\
%wrong sentences = 72 365 \\
%all sentences = 124 358 \\

%\textbf{WER 21,10 \% } \\

%\textbf{F1 words 78,90 \% } \\

%\textbf{F1 senctes 41,81 \% } \\

%\\

The WER (Word Error Rate), CER (Character Error Rate) and F1 metrics were used to evaluate the qualitative results of the models.
The WER/CER is defined as:

\begin{equation}
    WER/CER = \frac{I+D+S}{N} \times 100
\end{equation}

where all symbols denote the number of words in the transcription, respectively: S – substitutions (incorrect words/incorrect characters), D - deletions (removed words/removed characters), I – insertions (added words/added characters) and N – number of words/characters in the reference transcription. 
The F1 metric is defined as:

\begin{equation}
    F1 = \frac{2TP}{2TP+FN+FP}
\end{equation}

where TP - true positive, FN - false negative and FP - false positive.

\subsection{Multilingual LibriSpeech (MLS)}

In Table \ref{mls_results} the comparative results of three Scribe models are presented: ElevenLabs, Whisper, and Wav2Vec. They show that Whisper and ElevenLabs significantly outperform the Wav2Vec model (17.2\%). Furthermore, and remarkably, Whisper achieves a slightly lower WER (5.05\%) than the ElevenLabs model (6.77\%). This shows that with short context, colloquial vocabulary, simple phrases, and relatively good quality recordings, the Whisper model can perform comparable to the ElevenLabs model. 

The results in Table \ref{mls__results} confirm this hypothesis. The Whisper large version achieves slightly better results for WER and F1 for words, while the Scribe ElevenLabs model is by far the best in terms of F1 for sentences. It should be noted that the results of the Whisper medium model (6.84\% WER) are slightly worse than those of Scribe ElevenLabs (6.77\% WER). The Whisper small version differs significantly from the others in quality, showing significantly worse metrics (11.92\% WER). Whisper's small version still outperforms other ASR models for the MLS dataset.

\subsection{Mozilla Common Voice (MCV)}

For the Common Voice dataset, ElevenLabs' Scribe clearly performs best (Table \ref{cv_results} and \ref{cv__results}). Scribe ElevenLabs achieves an exceptionally low WER (3.29\%). Next in line are FastConformer and Whisper Large (5.99\% and 6.3\%). It is worth noting that the FastConformer model slightly outperforms Whipser large on this dataset. The remaining models achieve significantly worse results (from 9.8 and above). Table \ref{cv__results} shows a detailed comparison of the Scribe ElevenLabs models and the Whisper family of models. It shows a clear advantage of the Scribe model in all metrics. It should be noted that the Whisper Medium and Small models achieve results comparable to well-known models such as QuartzNet, EspNet, and Wav2Vec (about the 10\% and more). 

%\textbf{Without special characters}

%correct sentences = 202 \\
%wrong senteces = 260 \\
%all sentences = 462 \\ 
%all words = 17034 \\ 
%wrong words = 3038 \\
%correct words = 13996 \\

%\textbf{WER 17,80 \%}  \\

%\textbf{F1 words 82,20 \% } \\

%\textbf{F1 senteces 43,80\% } \\

%\textbf{With special characters}

%correct sentences = 178 \\
%wrong senteces = 283 \\
%all sentences = 462 \\
%all words = 17034 \\
%wrong words = 3182 \\
%correct words = 13852 \\

%\textbf{WER 18,60 \% } \\

%\textbf{F1 words 81,40 \% } \\

%\textbf{F1 sentences 36,80 \% } \\

\section{Results on medical interviews}
\label{sec:medical}

To facilitate quantitative analysis, the merged medical dataset integrates transcription-level metrics—word count, word and sentence error rates, and automatic alignment statistics. This data set also aggregates normalized and raw metrics (WER, CER) for multiple ASR models, enabling fine-grained comparison across medical domains. This structure provides a valuable foundation for further experiments on the domain adaptation and noise robustness of ASR systems in clinical use.

Evaluation of the quality of transcription of Polish medical records (Table~\ref{whisper_results}) demonstrates a considerable performance gap between generic multilingual models and a dedicated solution. The ElevenLabs model achieved a normalised Word Error Rate (WER) of 10.58\% and a Character Error Rate (CER) of 7.98\%, clearly outperforming Whisper, whose corresponding scores exceeded 40\% and 28\%, respectively. These results confirm that large commercial transcription engines fine-tuned for multilingual clinical speech maintain robustness even in low-resource languages such as Polish.

Beyond raw error rates, qualitative inspection revealed that Whisper frequently misinterprets domain-specific abbreviations, Latin expressions, and numerical data, while ElevenLabs preserved both lexical correctness and contextual coherence. This is crucial in medical practice, where transcription errors can propagate into patient documentation or automatic coding modules. Incorporating domain-adapted language models (e.g., Polish LLMs trained on medical corpora) could further improve recognition accuracy and note generation consistency.

In general, experiments show that precise transcription of doctor–patient conversations is achievable when combining a high-quality ASR front-end with specialized downstream processing. The presented framework forms the foundation for future work on end-to-end speech-driven medical documentation in Polish clinical settings.

\section{Conclusions}
\label{sec:conclusions}
The study presented in this paper compared several state-of-the-art ASR models for Polish languages with a focus on doctor–patient conversations used for automatic generation of medical notes. On public benchmarks, large Transformer-based models clearly dominate: among open-source systems, Whisper large offers the best overall performance and is a strong general-purpose baseline for Polish ASR.

However, experiments on real patient interviews show that the benchmark results do not transfer directly to the medical domain. ElevenLabs Scribe markedly outperforms Whisper on noisy, overlapping clinical speech with dense medical terminology, while generic multilingual models struggle with abbreviations, Latin terms and numerical data. This confirms that a high-quality, domain-adapted ASR is a critical prerequisite for reliable speech-driven automation of medical documentation, such as CORA AI, as it directly affects summarization and automatic coding.

% \subsection{Medical notes with good quality}

%\begin{table}
%\centering
%\begin{tabular}{|c|c|c|c|c|c|c|c|c|c|}
%\hline
%\textbf{} & \textbf{all words} & \textbf{wrong words} & \textbf{correct sentences} & \textbf{wrong sentences}  & \textbf{all sentences} & \textbf{WER}
%& \textbf{F1 words} & \textbf{F1 sentences}\\ \hline
%\textbf{ElevenLabs}     &    202     &  260  &      462          & 17 034 & 3 038 & 17.80 &   82.20  & 43.80           \\ \hline
%\textbf{Whisper}     &   178        &  283  &     462          &  17 034 &   3 182   &  18.60 &   81.40 & 36.80       \\ \hline
%\textbf{}     &         &                 &             & &    & &  \\ \hline

%\end{tabular}
%\caption{Comparison between different Whisper models and ElevenLabs}
%\label{whisper_results}
%\end{table}

% \subsection{Medical notes with poor quality}

% \subsection{Error analysis for medical data}

\begin{table}
\centering
\begin{tabular}{|c|c|c|}
\hline
\textbf{quality} & \textbf{ElevenLabs} & \textbf{Whisper} \\ \hline
 WER raw &  12.06   &     53.83    \\ \hline
 WER norm &   10.58  &     42.26           \\ \hline
 CER raw &   8.34  &    32.14     \\ \hline
 CER norm &  7.98   &    28.58     \\ \hline

\end{tabular}
\caption{ElevenLabs and Whisper comparison on medical interviews}
\label{whisper_results}
\end{table}

%\textcolor{red}{the ablation study - different chunks for LLM and different prompts, small, medium, large whisper, noise}

%\begin{table}
%\centering
%\begin{tabular}{|c|c|c|c|c|c|}
%\hline
%\textbf{} & \textbf{small Whisper} & \textbf{medium Whisper} & \textbf{large Whisper} & \textbf{ElevenLabs}  \\ \hline
%\textbf{Common Voice}     &         &   &                 &              \\ \hline
%\textbf{MLS LibriSpeech}     &           &    &                &               \\ \hline
%\textbf{}     &         &                 &             &      \\ \hline

%\end{tabular}
%\caption{Comparison between different Whisper models and ElevenLabs}
%\label{whisper_results}
%\end{table}

%\begin{table}
%\centering
%\begin{tabular}{|c|c|c|c|c|}
%\hline
%\textbf{} & \textbf{chunk - } & \textbf{chunk - } & \textbf{chunk - }  \\ \hline
%\textbf{Common Voice}     &         &                   &               \\ \hline
%\textbf{Vox Populi}     &           &                     &              \\ \hline
%\textbf{}     &         &                             &       \\ \hline

%\end{tabular}
%\caption{Comparison of chunk size and its influence on LLM accuracy}
%\label{whisper_results}
%\end{table}

\bibliographystyle{unsrt}  


\end{document}